\documentclass[12pt,nohyper,letterpaper]{JHEP3}  

\usepackage{amsmath,amssymb,amsfonts,amsxtra,makeidx,graphics,graphicx,amsthm,epstopdf,epsfig}

\newcommand{\be}{\begin{equation}}
\newcommand{\ee}{\end{equation}}
\newcommand{\beq}{\begin{equation}}
\newcommand{\eeq}{\end{equation}}
\newcommand{\ba}{\begin{array}}
\newcommand{\ea}{\end{array}}
\newcommand{\bea}{\begin{eqnarray}}
\newcommand{\eea}{\end{eqnarray}}
\newcommand{\ben}{\begin{enumerate}}
\newcommand{\een}{\end{enumerate}}
\newcommand{\bean}{\begin{eqnarray*}}
\newcommand{\eean}{\end{eqnarray*}}
\newcommand{\eref}[1]{(\ref{#1})}

\newcommand{\nn}{\nonumber}

\newcommand{\BC}{\mathbb{C}}
\newcommand{\BR}{\mathbb{R}}
\newcommand{\BZ}{\mathbb{Z}}
\newcommand{\comment}[1]{}

\newcommand{\CM}{{\cal M}}

\newcommand{\CN}{{\cal N}}

\newcommand{\IP}{\mathbb{P}}

\newcommand{\CP}{\mathbb P}

\newcommand{\PE}{\mathrm{PE}}
\newcommand{\PL}{\mathrm{PL}}

\newcommand{\setall}{\setcounter{equation}{0}
        \setcounter{theorem}{0}}

 \title{Partition Functions for Membrane Theories}
\author{Amihay Hanany and Noppadol Mekareeya \\
Theoretical Physics Group, The Blackett Laboratory \\
Imperial College London, Prince Consort Road\\ 
London,  SW7 2AZ,  UK \\
{\tt a.hanany, n.mekareeya07@imperial.ac.uk}}

\author{Alberto Zaffaroni \\
Universit\`a di Milano-Bicocca and INFN \\
sezione di Milano-Bicocca, Piazza della Scienza, 3;\\
I-20126 Milano, Italy \\
{\tt alberto.zaffaroni@mib.infn.it}}

\abstract{Partition functions for M2-brane theories in various backgrounds are 
computed. We consider in particular configurations of membranes at orbifold
singularities preserving $\CN=5$ or $\CN=6$ supersymmetry. The worldvolume
membrane theory for some of these configurations has been recently constructed
in terms of $\CN=6$ Chern-Simons theories. 
The detailed structure of the partition functions as well as their transformation rules under the R-symmetry are explicitly computed using the Plethystic Programme.}

\begin{document}
\pagestyle{plain}
\setcounter{page}{1}
\newcounter{bean}
\baselineskip16pt

\section{Introduction}
Partition functions for BPS operators in supersymmetric field theories are rather interesting objects
in many respects. They share information about  the structure of the moduli space of vacua
and the effective number of degrees of freedom in the system. The computation of such generating functions is generically a very hard problem but it can be simplified in particular
circumstances. The partition functions for chiral
operators in four dimensional supersymmetric gauge theories   have been extensively studied in the 
past years, ranging from SQCD \cite{ pouliot, romelsberger,Hanany:2008kn, Gray, hanany} to quiver gauge theories living on branes at singularities \cite{Kinney:2005ej,Martelli:2006vh,BFHH,  Butti:2006au,Hanany:2006uc, feng,Forcella:2007wk,Butti:2007jv, Forcella:2007ps, Forcella:2008bb}.
The latter in particular are superconformal gauge theories and have an  $AdS_5\times H$ dual \cite{Klebanov:1998hh, Acharya:1998db, Morrison:1998cs}.
In this case, information from the field theory and from the holographic dual can be combined to give a better understanding of the superconformal theory. For the case of branes at Calabi-Yau
singularities, the combination of the Plethystic Programme with algebraic tools in complex geometry
allows to write quite explicit formulae for the partition functions. 

It is a natural and interesting
direction to try and extend these results to other dimensions. In particular, the case of three dimensions,
where the superconformal zoo is very large, is a natural choice. Most supersymmetric
Yang-Mills theories flow in the IR to a superconformal fixed point in three dimensions. However, for theories with an $AdS_4\times H$ dual, it is very difficult to write the corresponding UV Yang-Mills
theory\footnote{We note in passing that the analogous problem in four dimensions is solved, at least for toric Calabi-Yaus, by the dimer technology \cite{Hanany:2005ve, Franco:2007ii, Franco:2005rj}. In three dimensions there is a proposal  based on crystals \cite{Lee:2007kv}.} and little is known about the explicit description of the interacting superconformal theory which is assumed to be a theory of membranes. 

In this paper we will consider theories of membranes living at singularities
$\mathbb{R}^8/\Gamma$ and preserving $\CN=5$ or $\CN=6$ supersymmetry. The dual
theory is $AdS_4\times S^7/\Gamma$. Here $\Gamma$ is any of the discrete 
subgroups of $SU(2)$ and it acts freely on $S^7$.

One of the motivation for this analysis is the fact that a superconformal Chern-Simons theory with $\CN=6$ supersymmetry and moduli space $\mathbb{R}^8/\BZ_k$
has been recently constructed. In fact there was recently much activity in the study of superconformal Chern-Simons theories in three dimensions with large amount of supersymmetry in the attempt of constructing theories for M2-branes.
A consistent theory with $\CN=8$ supersymmetry has been constructed with gauge group $SU(2)\times SU(2)$ in \cite{Bagger:2007vi} and interpreted
as the theory of two M2-branes on an orbifold of flat space for some value of the Chern-Simons parameter \cite{Distler:2008mk,Lambert:2008et}. Attempts to extend this construction to $N$ branes and $SU(N)$ gauge groups keeping manifest $\CN=8$ supersymmetry  faced intrinsic difficulties in the theory of three Lie algebras.
The only available candidates at the moment contain ghosts \cite{ghost1, ghost2, ghost3} \footnote{Ghosts can be consistently eliminated but it seems that the resulting theory is $\CN=8$ SYM and not its infrared limit \cite{ghost4, ghost5,Ezhuthachan:2008ch}.}. However, more recently, a consistent
theory with $U(N)\times U(N)$ gauge group and bifundamental fields which has only a manifest $\CN=6$ supersymmetry  has been constructed in
\cite{Aharony:2008ug}. The theory has two parameters,
the Chern-Simons parameter $k$  and the number of colors $N$. Based on the the analysis
of the moduli space, the spectrum of chiral operators and a brane construction, this theory has been proposed as the superconformal theory living on $N$ M2-branes at the orbifold singularity $\mathbb{R}^8/\BZ_k$.
Further evidence of this fact was given in \cite{Benna:2008zy,Bhattacharya:2008bj}.
The theory has a dual description as string theory on $AdS_4\times S^7/\BZ_k$.
In particular, for Chern-Simons level $k=1$ we recover the maximally supersymmetric theory
of M2-branes in flat space; only an $\CN =6$ supersymmetry is however manifest in the Lagrangian.

It is then a natural question to write partition functions 
for these theories. Of course, many things are known about the chiral 
spectrum of M2-branes in flat space.
A partition function for $1/8$ BPS operators was written in \cite{Bhattacharyya:2007sa}, for example. For theories with such amount of supersymmetry, it is natural and convenient to write down partition functions which respect the R-symmetry of the superconformal theory, which is $Spin(8)$ for $\CN=8$. 
This can be very efficiently done in the case of one brane.  

In fact, it is known from the  $AdS_4\times S^7$ dual description that there is precisely one single trace chiral multiplet for each symmetric traceless representation of $Spin(8)$.  Making use of this information,
%\footnote{Paper \cite{Aharony:2008ug} claims that the spectrum of $AdS_4\times S^7$ is in agreement with the spectrum of $\CN=6$ the Chern-Simons theory.}, 
we show how to write the partition function of one M2-brane in flat space and expand it in terms of $Spin(8)$ representations\footnote{We remark that, in obtaining the partition function, the Bagger--Lambert or a related theory is not directly applied.  Rather, we make use of the expected structure of the moduli space of M2-branes, \emph{i.e.} $AdS_4\times S^7/\Gamma$.}. 
The supersymmetric partition function
on $\mathbb{R}^8/\BZ_k$ is then obtained by using the discrete Molien formula \eref{dismo}
and expanded in terms of representations of the R-symmetry group $Spin(6)$.
We will also write the partition function for the theory of an M2-brane on
$\mathbb{R}^8/\hat{D}_{k+2}$ and $\mathbb{R}^8/\hat{E}_{n}$ (with $n=6,~7,~8$), configurations that preserves $\CN=5$ supersymmetry and
has $Sp(2)$ R-symmetry.   We emphasise that the partition functions precisely count chiral multiplets whose lowest compontent is a scalar field.

We can apply the Plethystic Programme in order to get information on the 
moduli space for higher $N$, which is ${\rm Sym}(\mathbb{R}^8/\Gamma)^N$, where $\Gamma$ can be the abelian group $\mathbb{Z}_k$ or any of the non-abelian discrete subgroups of $SU(2)$ associated with the affine Dynkin diagrams of $D_{k+2}, E_6, E_7, E_8$ \cite{BFHH, Hanany:1998sd}. Below we write a quick reminder.

\paragraph{The Plethystic Programme: A Recapitulation.} 
%From the supersymmetric partition function $g_1(t)$ for one brane
%on $\mathbb{R}^8$, choosing carefully the object we want to count and introducing
%chemical potentials (fugacities) $t$ for their symmetries.
%we can obtain the partition function at finite $N$ as follows.  
Let us define the {\bf plethystic exponential} of a  multi-variable function $ g (t_1 , \ldots , t_n )$ that vanishes at the origin, $g (0,\ldots, 0) = 0$, to be 
\begin{equation}
\label{PE}
\PE [g (t_1 , \ldots , t_n )] := \exp \left ( \sum_{r=1}^\infty \frac{g(t_1^r,\ldots,t_n^r)}{r}\right ) \, .
\end{equation}
In the same way as mentioned in \cite{BFHH, feng}, the generating function $g_N$ at finite $N$ is found by the series expansion of the $\nu$ - inserted plethystic exponential as 
\begin{equation}
\PE[\nu g_1(t_1,\ldots,t_n)] =\exp \left(\sum_{r=1}^\infty \frac{\nu^r g_1(t_1^r,\ldots,t_n^r)}{r} \right )= \sum_{N=0}^\infty g_N(t_1,\ldots,t_n) \nu^N \, .
\end{equation}
Information about the generators of the moduli space and the relations they satisfy can be computed by using the {\bf plethystic logarithm}, which is the inverse function of the plethystic exponential. Using the M\"obius function $\mu(r)$ we define
\begin{equation}
\label{PL}
\PL [g (t_1 , \ldots , t_n )] := \sum_{r=1}^\infty \frac{\mu(r) \log g(t_1^r,\ldots,t_n^r)}{r} \, .
\end{equation}
The significance of the series expansion of the plethystic logarithm is stated in \cite{BFHH, feng}: \emph{the first terms with plus sign give the basic generators while the first terms with the minus sign give the constraints between these basic generators.}  If the formula (\ref{PL}) is an infinite series of terms with plus and minus signs, then the moduli space is not a complete intersection and the constraints in the chiral ring are not trivially generated by relations between the basic generators, but receive stepwise corrections at higher degree. These are the so-called higher syzygies. 

These partition functions can be decomposed into representations of the relevant
R-symmetry group, $Spin(8),Spin(6)$ and $Spin(5)$ for $\CN= 8,6$ and $5$ supersymmetry
respectively. A word of caution is necessary. These partition functions count some
gauge invariant multitrace operators.
However, since the product of short multiplets of $\CN= 8, 6$ and 5 supersymmetry
may contain operators that are not protected, the partition functions are not necessarily
counting short operators except for $N=1$. They should be better intended
as partition functions counting real functions on the moduli space for $N$
branes. It would be interesting to investigate further the properties
of these partition functions and to seek for a dual interpretation for them.

\noindent \paragraph{Notation for representations.} In this paper, we shall represent an irreducible representation of a group $G$ by its highest weight $[a_1, \ldots, a_r]$, where $r = \mathrm{rank}~G$.  In order to avoid cluttered notation, we shall also slightly abuse terminology by referring to each character by its corresponding representation.

We next proceed with a detailed study of this class of theories.

\section{The Theory of $N=1$ and $k=1$} \label{11}  \setall
This theory has ${\cal N}=8$ supersymmetry in 2+1 dimensions and therefore all protected operators appear in irreducible representations of the R-symmetry group, $Spin(8)$. There is an additional quantum number which counts the number of scalar fields. This quantum number can be taken to be the conformal dimension of the corresponding operators, measured in units of $1/2$. Its corresponding fugacity is denoted by $t$. The moduli space is $\BR^8$ and the scalars transform in the $[1,0,0,0]$ representation of $Spin(8)$. It should be noted that the fugacity $t$ represents a real degree of freedom and not a complex degree of freedom. As a result, the dimension of the moduli space is real and not complex.
We therefore write down the first partition function for the set of theories $g_{N,\BZ_k}$ (below $\BZ_1$ means the trivial action),
\begin{equation}
g_{1,\BZ_1} (t, y_1,y_2,y_3, y_4; \BR^8 ) = \PE [ [1,0,0,0] t ].
\end{equation}
Here and below the notation $[1,0,0,0]$ is taken to be the character of the representation with these highest weights. To be concrete we can choose four complex fugacities $y_1, y_2, y_3, y_4$ such that
\begin{equation}
[1,0,0,0] = \frac{y_1}{y_2}+y_1+\frac{y_3
   y_4}{y_2}+\frac{y_4}{y_3}+\frac{y_3}{y_4}+\frac{y_2}{y_3
   y_4}+\frac{y_2}{y_1}+\frac{1}{y_1} .
\end{equation}
Using this, an explicit expression for $g_{1,\BZ_1}$ after evaluating the $\PE$, as defined in \eref{PE}, takes the form
\beq \ba{rcl}
g_{1, \BZ_1} (t, y_1,y_2,y_3, y_4; \BR^8 ) = \frac{1}{\left(1-\frac{t}{y_1}\right) \left(1-t y_1\right) \left(1-\frac{t y_1}{y_2}\right) \left(1-\frac{t y_2}{y_1}\right) \left(1-\frac{t y_2}{y_3 y_4}\right) \left(1-\frac{t y_3}{y_4}\right) \left(1-\frac{t y_4}{y_3}\right) \left(1-\frac{t y_3 y_4}{y_2}\right)} .
\label{explicitg11}
\ea
\eeq
The partition function $g_{1,\BZ_1}$ has an expansion in terms of characters of $Spin(8)$ as
\begin{equation}
g_{1,\BZ_1} (t, y_1,y_2,y_3, y_4; \BR^8 ) = 1 + [1,0,0,0] t + ( [2,0,0,0] + [0,0,0,0] ) t^2 + \ldots
\end{equation}
When we set all the chemical potentials of the $Spin(8)$ symmetry to zero this function takes the form
\begin{equation}
g_{1,\BZ_1} (t,1,1,1,1; \BR^8 ) = \frac{1}{(1-t)^8} = 1+8t + 36 t^2 +\ldots .
\end{equation}
We first note that this function has a pole of order 8 at $t=1$ which indicates that the real dimension of the moduli space is 8.

\paragraph{Operators on $S^7$.} This partition function turns out to count operators which are not protected by supersymmetry, the simplest one being $\mathrm{Tr} ( \phi_i \phi_i ) $, which is represented by the singlet term in the expansion at order $t^2$. To cure this we recall that the protected operators are actually in one to one correspondence with harmonic functions on $S^7$ (see, \emph{e.g.}, \cite{Biran:1983iy} for Kaluza--Klein modes on $S^7$)\footnote{References \cite{Biran:1983iy, Nilsson:1984bj} contain an 11 dimensional supergravity argument from which the zero modes for a single M2-brane can be read off.}, and the partition function should reflect this condition. It is easily given by a relation which is quadratic in the basic fields and a singlet of $Spin(8)$.
We therefore write a partition function for all harmonic functions on $S^7$,
\begin{equation}
g_{1,\BZ_1} (t, y_1,y_2,y_3, y_4; S^7 ) = (1-t^2) \PE \left [ [1,0,0,0] t \right ].
\end{equation}
This partition function has a nice expansion in terms of characters:
\begin{equation} \label{11so8}
g_{1,\BZ_1} (t, y_1,y_2,y_3, y_4; S^7 ) = \sum_{n=0}^\infty [n,0,0,0] t^n,
\end{equation}
which indeed reflects the well known fact that harmonic functions at level $n$ on $S^7$ transform as precisely one copy of the $[n,0,0,0]$ representation of $Spin(8)$. Correspondingly, the ${\cal N}=8$ theory for one M2-brane has a set of protected operators at level $n$ which transform under precisely one copy of the representation $[n,0,0,0]$ of $Spin(8)$.
We can further set all $Spin(8)$ chemical potentials to zero and get the expressions
\begin{equation}
g_{1,\BZ_1} (t, 1,1,1,1; S^7 ) = \frac{1-t^2}{(1-t)^8} = \frac{1+t}{(1-t)^7} = \sum_{n=0}^\infty \frac{(n+3)}{3} {n+5 \choose 5} t^n,
\end{equation}
The first form indicates that there are 8 generators for $S^7$ which are subject to 1 relation of order 2. This relation sets the radius of the $S^7$ to a constant value. The second form indicates that the real dimension of the moduli space is 7. The last form gives the dimensions of the irreducible representations $[n,0,0,0]$ of $Spin(8)$. For reference we quote here the general dimension formula \cite{FH} for a generic $Spin(8)$ representation of highest weight $[n_1, n_2, n_3, n_4]$:
\begin{equation}
\dim [n_1, n_2, n_3, n_4] = \frac{ \{1\} \{2\} \{3\} \{4\} \{12\}\{23\}\{24\} \{123\}\{234\}\{124\} \{1234\}\{12234\}} {4320} ,
\end{equation}
with $\{i\} = n_i +1$, $\{ij\} = n_i + n_j +2$, $\{ijk\} = n_i + n_j + n_k + 3$, \emph{etc.}

\paragraph{Decomposing $Spin(8)$ into $SU(4) \times U(1)$.} For applications with higher CS level $k$, we will now rewrite the generating functions in terms of irreducible representations of $SU(4)$, the R-symmetry for ${\cal N}=6$ supersymmetry in 2+1 dimensions.
For this purpose we introduce the fugacity $b$ for the baryonic number, and decompose the 8 dimensional representation of $SO(8)$ into two irreducible representations of $SU(4)$:
\begin{equation}
[1,0,0,0] t = [1,0,0] t_1 + [0,0,1] t_2
\end{equation}
with the usual relation as borrowed from the conifold partition functions, $t_1=t b, t_2 = t/b$ (see \emph{e.g.}, \cite{Forcella:2007wk}). Here $t_1$ is taken to count the degree of holomorphic functions on $\BC^4$ and $t_2$ counts the degree of anti-holomorphic functions on $\BC^4$. Explicit expressions for the characters of the $SU(4)$ representations can be taken to be with 3 complex fugacities, $z_1, z_2, z_3$ in the form,
\begin{equation}
[1,0,0] = z_1 + \frac{z_2}{z_1} + \frac{z_3}{z_2} + \frac{1}{z_3}, \qquad [0,0,1] = \frac{1}{z_1} + \frac{z_1}{z_2} + \frac{z_2}{z_3} + {z_3}
\end{equation}
The generating function $g_{1,\BZ_1}$ takes the form
\begin{equation}
g_{1,\BZ_1} (t_1, t_2, z_1,z_2,z_3; S^7 ) = (1-t_1 t_2) \PE [ [1,0,0] t_1 + [0,0,1] t_2 ].
\end{equation}
This function has a nice expansion in terms of irreducible representations of $SU(4)$,
\begin{equation} \label{11su4}
g_{1,\BZ_1} (t_1, t_2, z_1,z_2,z_3; S^7 ) = \sum_{n=0}^\infty \sum_{m=0}^\infty [n,0,m] t_1^n t_2^m = \sum_{n=0}^\infty \sum_{m=0}^\infty [n,0,m] b^{n-m} t^{n+m}.
\end{equation}
Comparing \eref{11so8} with \eref{11su4}, we find the following decomposition:
\beq \label{decomp}
[n,0,0,0]_{Spin(8)} \rightarrow \sum_{r=0}^n b^{n-2r} [n-r,0,r]_{SU(4)}~.
\eeq

\paragraph{A non-trivial check.}The dimension formula \cite{FH} for a generic representation of $SU(4)$ of highest weights $[n_1, n_2, n_3]$ is
\bea
\dim [n_1, n_2, n_3] &=& \frac{ \{1\} \{2\} \{3\} \{12\} \{23\} \{123\}} {12} \nn \\
&=& \frac{(n_1+1) (n_2+1)(n_3+1) (n_1+n_2+2) (n_2+n_3+2) (n_1+n_2+n_3+3)} {12}~.\nn\\
\eea
This can be used in checking the various relations quoted above and below.

%%%%%%%%%%%%%%%%%
\section{$\BZ_k$ Orbifold Actions on the $N=1$ Theory} \setall

%%%%%%%%%%%%%%%%%
\subsection{The Case of $k =2$}
We next turn to the $k=2$ theories. The R-symmetry is still $Spin(8)$ and we can still count operators using representations of $Spin(8)$. The new ingredient is an orbifold projection on the variable $t$.
Under this orbifold action $t\rightarrow -t$ and we need to sum over both sectors, with $t$ and with $-t$.
The resulting generating function gets a simple form, restricting to even powers of $t$,
\begin{equation} \label{12char}
g_{1,\BZ_2} (t, y_1,y_2,y_3, y_4; S^7 ) = \sum_{n=0}^\infty [2n,0,0,0] t^{2n} ~.
\end{equation}
Setting the $Spin(8)$ chemical potentials to zero, we find
\begin{equation}
g_{1,\BZ_2} (t, 1,1,1,1; S^7 ) = \frac{1+28 t^2+70 t^4+28 t^6+t^8}{\left(1-t^2\right)^7}~ ,
\end{equation}
suitable for a moduli space of real dimension 7.

The plethystic logarithm of the generating function $g_{1,\BZ_2}$ is
\bea
\PL \left[ g_{1,\BZ_2} (t, y_1,y_2,y_3, y_4; S^7 ) \right] &=&[2,0,0,0] t^2 - \left( [2,0,0,0] + [0,2,0,0] +[0,0,0,0] \right) t^4 + \ldots~, \nn\\
\PL \left[ g_{1,\BZ_2} (t, 1,1,1, 1; S^7 ) \right] &=& 35t^2 - 336t^4+5376 t^6-101856 t^8 \ldots~.
\eea
This indicates that there are 35 basic generators transforming in the $SO(8)$ representation $[2,0,0,0]$ at order $t^2$, and there are 336 basic relations transforming in the representations $[2,0,0,0] + [0,2,0,0] +[0,0,0,0]$ at order $t^4$. We note that in this case the moduli space is not a complete intersection, since the plethystic logarithm is an infinite series.
%In terms of $SU(4)$ irreducible representations, this expression can be written using the decomposition \eref{decomp} as
%\beq
%g_{1,\BZ_2} (t, b ,z_1,z_2,z_3; S^7 ) = \sum_{n=0}^\infty \sum_{r=0}^n b^{n-2r} [2n-r,0,r] t^{2n}~. 
%\eeq
 
%The plethystic logarithm of this expression is
%\beq
%PL \left[g_{1,\BZ_2} (t, 1,1,1, 1; S^7 ) \right] = 35 t^2-336 t^4+5376 t^6-101856 t^8 + \ldots~.
%\eeq

\subsection{The Case of Higher $k$}
For higher values of $k$ the orbifold action does not commute with the $Spin(8)$ R-symmetry group and breaks it to $SU(4)$ with an action on the baryonic charge. The $\BZ_k$ orbifold acts on the fugacity $b$ by $b\rightarrow w b$, with $w^k=1$ and we need to sum over all contributions. The result is the following discrete Molien formula (\emph{c.f.} Equation (3.1) of \cite{BFHH}):
\begin{equation} \label{1k}
g_{1,\BZ_k} (t, b, z_1, z_2, z_3) = \frac{1}{k} \sum_{j=0}^{k-1} g_{1,\BZ_1} (t, w^j b, z_1, z_2, z_3) 
\end{equation}
It is now useful to recall \eref{11su4} and realize that only terms with $n-m = 0\mod k$ survive the projection. We can therefore write an expression for $g_{1,\BZ_k}$ as follows:
\bea 
g_{1,\BZ_k} (t, b, z_1, z_2, z_3; S^7) &=& \sum_{n_1=0}^\infty \sum_{n_2=0}^\infty \sum_{r=0}^{k-1} [kn_1 +r, 0, kn_2+r] t_1^{kn_1+r} t_2^{kn_2+r} \nn \\
&=& \sum_{n_1=0}^\infty \sum_{n_2=0}^\infty \sum_{r=0}^{k-1} [kn_1 +r, 0, kn_2+r]  b^{k(n_1-n_2)} t^{k(n_1+n_2) +2r}~. \nn \\
\label{g1k}
%g_{1,\BZ_k} (t, b, z_1, z_2, z_3) = \sum_{r=0}^\infty \sum_{s=-\lfloor\frac{r}{k}\rfloor}^\infty [r + k s,0,r] t_1^{r+k s} t_2^{r} = \sum_{r=0}^\infty \sum_{s=-\lfloor\frac{r}{k}\rfloor}^\infty [r+ k s,0,r] b^{k s} t^{2 r + k s} ,
\eea
%where the limits on the values of $r$ and $s$ are such that only positive integers are entering into the highest weight representation.  
We shall see in examples below that, for an arbitrary CS level $k$, the generators are in the representations  $[1,0,1], [k,0,0]$ and $[0,0,k]$. This is consistent with the analysis of chiral operators performed in \cite{Aharony:2008ug} for the $\CN=6$ CS theory. 

\paragraph{An Example of $k=2$.} As a check, we can recover the previous results for $k=2$. 
Formula \eref{g1k} gives 
\beq \label{g12}
g_{1,\BZ_2} (t, b, z_1, z_2, z_3; S^7) = 1+\left(b^2[2,0,0] + [1,0,1] + \frac{1}{b^2}[0,0,2] \right) t^2 + \ldots~. 
\eeq
Setting $b=z_1=z_2=z_3=1$, we have the unrefined partition function
\bea \label{g12}
g_{1,\BZ_2} (t, 1, 1, 1, 1; S^7) &=& \frac{1+28 t^2+70 t^4+28 t^6+t^8}{\left(1-t^2\right)^7} \nn \\
&=& 1+35 t^2+294 t^4+1386 t^6+4719 t^8+13013 t^{10}+\ldots~.
\eea
The plethystic logarithm of this expression is
\bea
\PL \left[ g_{1,\BZ_2} (t, 1, 1, 1, 1; S^7) \right] = 35 t^2 - 336 t^4 + 5376 t^6 - 101856 t^8+\ldots~.
\eea
Observe that the coefficient 35 of $t^2$ in the plethystic logarithm is simply the dimension of the $SU(4)$ representations $[1,0,1] + [2,0,0] + [0,0,2]$ in the second term of \eref{g12}. This indicates that the generators transform in the representations  $[1,0,1]$, $[2,0,0]$ and $[0,0,2]$, which is indeed the decomposition of the $[2,0,0,0]$ representation of $Spin(8)$.   

\paragraph{An Example of $k=3$.}  The unrefined partition function is
\bea \label{g13}
g_{1,\BZ_3} (t, 1, 1, 1, 1; S^7) &=& \frac{1-3 t+18 t^2-10 t^3+21 t^4+21 t^5-10 t^6+18 t^7-3 t^8+t^9}{(1 - t)^7 (1 + t + t^2)^4} \nn \\
&=& 1+15 t^2+40 t^3+84 t^4+240 t^5+468 t^6+840 t^7+\ldots~.
\eea 
The plethystic logarithm of this expression is
\bea
\PL \left[ g_{1,\BZ_3} (t, 1, 1, 1, 1; S^7) \right] = 15 t^2+40 t^3-36 t^4-360 t^5-492 t^6+2880 t^7+\ldots~.
\eea
We note that the coefficient 15 of $t^2$ is the dimension of the representation $[1,0,1]$, and the coefficient $40$ of $t^3$ is the dimension of $[3,0,0] + [0,0,3]$.  This indicates that the generators transform representations $[1,0,1]$, $[3,0,0]$ and $[0,0,3]$.   

\paragraph{An Example of $k=4$.}  The unrefined partition function is
\bea \label{g14}
g_{1,\BZ_4} (t, 1, 1, 1, 1; S^7) &=&\frac{1+12 t^2+108 t^4+212 t^6+358 t^8+212 t^{10}+108 t^{12}+12 t^{14}+t^{16}}{\left(1-t^2\right)^7 \left(1+t^2\right)^4} \nn \\
&=& 1+15 t^2+154 t^4+678 t^6++2387 t^8 + 6461 t^{10}\ldots~.
\eea 
The plethystic logarithm of this expression is
\bea
\PL \left[ g_{1,\BZ_4} (t, 1, 1, 1, 1; S^7) \right] = 15 t^2+34 t^4-512 t^6+2332 t^8+\ldots~.
\eea
The coefficient 15 of $t^2$ is the dimension of the representation $[1,0,1]$.  The coefficient $34$ of $t^4$ is the dimension of $[4,0,0]+[0,0,4] - \left( [0, 2, 0] + [1, 0, 1] + [0,0,0] \right)$.  We note that the correction $[0, 2, 0] + [1, 0, 1] +[0,0,0]$, which is contained in the decomposition of $\mathrm{Sym}^2[1,0,1]$, is simply the relation at order $t^4$.  Therefore, the generators transform under the representations $[1,0,1]$, $[4,0,0]$ and $[0,0,4]$.   

\paragraph{An Example of $k=5$.}  The power series of the unrefined partition function is
\bea
g_{1,\BZ_5} (t, 1, 1, 1, 1; S^7) = 1+15 t^2+84 t^4+112 t^5+300 t^6+560 t^7+825 t^8 +\ldots~.
\eea 
The plethystic logarithm of this expression is
\bea
\PL \left[ g_{1,\BZ_5} (t, 1, 1, 1, 1; S^7) \right] = 15 t^2-36 t^4+112 t^5+160 t^6+\ldots~.
\eea
The coefficient 15 of $t^2$ is the dimension of the representation $[1,0,1]$.  The coefficient $-36$ of $t^4$ indicates that there are relations transforming in the representation $[0, 2, 0] + [1, 0, 1] +[0,0,0]$ at order $t^4$, as before. The coefficient 112 of $t^5$ is the dimension of $[5,0,0]+[0,0,5]$. Therefore, the generators transform representations $[1,0,1]$, $[5,0,0]$ and $[0,0,5]$. 

\paragraph{A general formula.}  The general unrefined partition function which can be obtained from \eref{1k} is 
\bea 
g_{1,\BZ_k} (t, 1, 1, 1, 1; S^7) &=&  \frac{1}{3 (1 - t^2)^6 (1 - t^k)^4} \times \Big[ 3 + 27 t^2 + 27 t^4 + 3 t^6 + \big( -6 + 11 k + 6 k^2 + k^3 \nn \\
&& - 54 t^2 + 27 k t^2 - 6 k^2 t^2 - 3 k^3 t^2 - 54 t^4 - 27 k t^4 - 6 k^2 t^4 + 3 k^3 t^4 - 6 t^6 \nn\\
&& - 11 k t^6 + 6 k^2 t^6 - k^3 t^6 \big) t^k  + \big(-22 k + 4 k^3 - 54 k t^2 - 12 k^3 t^2 + 54 k t^4  \nn \\
&& + 12 k^3 t^4 +22 k t^6 - 4 k^3 t^6 \big) t^{2 k} + \big(6 + 11 k - 6 k^2 + k^3 + 54 t^2 + 27 k t^2 \nn \\
&& + 6 k^2 t^2 - 3 k^3 t^2 + 54 t^4 - 27 k t^4 + 6 k^2 t^4 + 3 k^3 t^4 + 6t^6 - 11 k t^6 - 6 k^2 t^6  \nn \\
&& - k^3 t^6 \big) t^{3 k} + \big(-3 - 27 t^2 - 27 t^4 - 3 t^6 \big) t^{4 k} \Big]~. \label{unrefzk}
\eea

\subsection{The $k\rightarrow\infty$ Limit: Restriction to the Zero Baryonic Subspace} 
In the limit where $k$ goes to infinity, all states with non zero
baryonic charge disappear from the spectrum.  We obtain a partition function 
which counts real functions on $\IP^3$,
\begin{equation} \label{zerobar}
g_{1,\BZ_k} (t, z_1,z_2,z_3; \IP^3 ) = \sum_{n=0}^\infty [n,0,n] t^{2n}~,
\end{equation}
where the $SU(4)$ representation $[n,0,n]$ can be interpreted as the partition function for $\CN =6$ chiral multiplets in the Kaluza-Klein (KK) compactification on $\CP^3$.
It is well known indeed that the KK chiral multiplets for $\mathrm{AdS}_4 \times \CP^3$ fall in $[n, 0, n]$ representations \cite{Nilsson:1984bj}.

When restricted to zero $SU(4)$ chemical potentials, we get
\begin{equation}
g_{1,\BZ_k} (t, 1,1,1; \IP^3 ) = \frac{1+9t^2 + 9 t^4 +t^6}{(1-t^2)^6} = \sum_{n=0}^\infty \frac{(n+1)^2 (n+2)^2 (2 n+3)}{12} t^{2n}~,
\end{equation}
where we note that this formula agrees with $\eref{unrefzk}$ in the limit $k \rightarrow \infty$.

It is obvious from the order of the pole that we are dealing with a six dimensional manifold.
This is explained by the fact that $\BZ_k$ acts by reducing by a factor of $k$ the length of a circle in $S^7$. In the limit $k\rightarrow\infty$,
$S^7$ becomes $\mathbb{P}^3$ and, correspondingly, M-theory is reduced to Type IIA.
The above partition function  then characterises the protected Type IIA configurations on 
$AdS_4\times \IP^3$ \cite{Aharony:2008ug}.
%which characterises the relevant Type IIA configuration on $AdS_4\times \IP^3$.
We point out that this partition function is palindromic even though it is not a CY manifold.

%%%%%%%%%%%%%%%%%%%%%%%%%%%%%%
\section{Non-Abelian Orbifold Actions on the $N=1$ Theory} \setall
We now consider the orbifold actions of the binary dihedral,\footnote{In this paper, we shall denote the binary dihedral group of order $4k$ by $\hat{D}_{k+2}$.} tetrahedral, octahedral and icosahedral discrete subgroups $\Gamma$ of $SU(2)$ associated to the affine lie algebras $\hat{D}_{k+2},~\hat{E}_{6},~\hat{E}_{7},~\hat{E}_{8}$, whose projections break the $Spin(8)$ R-symmetry group into $Sp(2)$ and preserves $\CN=5$ supersymmetry\footnote{$\mathbb{C}^4/\Gamma$, with $\Gamma$ discrete subgroup of $SU(2)$ acting diagonally on two copies of $\mathbb{C}^2$, is obviously a Calabi-Yau cone on $S^7/\Gamma$. It preserves $\CN=6$ supersymmetry for Abelian $\Gamma$ and $\CN=5$ for dihedral and exceptional $\Gamma$ \cite{Morrison:1998cs} , as can be checked by the action 
on spinors. Notice however that in the text we adopted a real notation
which is related to the Calabi-Yau complex coordinates by a change of complex structure.}. 
We note that the membrane theory on $\mathbb{R}^8/\Gamma$ has a dual $AdS_4\times S^7/\Gamma$. 

\paragraph{Discrete Molien formula.} The partition function for $S^7/\Gamma$ depending on the parameter $t$ can be easily computed by the following discrete Molien formula (\emph{c.f.} Equation (3.1) of \cite{BFHH}):
\begin{equation} \label{dismo}
g_{1,\Gamma} (t) = \frac{1}{|\Gamma|} \sum_{\gamma\in \Gamma} \frac{1-t^2}{\mathrm{det}(I_{8\times 8} - t \gamma)}~,
\end{equation}
where the determinant is taken over the $8 \times 8$ matrix representation of the group elements.

\paragraph{Decomposing $SU(4)$ into $Sp(2)$.} Since the $SU(4)$ R-symmetry is broken into $Sp(2)$, we will need to expand partition functions in terms of irreducible representations of $Sp(2)$ instead of $SU(4)$.  We shall quote here the relevant decomposition formula (setting the fuacities $z_i$ of $SU(4)$ to the fugacities $x_i$ of $Sp(2)$ to be $z_1 = x_1, z_2 = x_2, z_3 = x_1$; this action is like a ``folding" of the representation similar to the action of an orientifold plane.):
\beq
[m, 0, n]_{SU(4)} \rightarrow \sum_{a=0}^{\mathrm{min}\{ m,n \}} [m+n-2a, a]_{Sp(2)}~.
\eeq
Therefore, we can rewrite \eref{g1k} in terms of $Sp(2)$ irreducible representations, setting $b=1$ since the baryonic charge is not conserved with non-Abelian orbifold projections,
\bea \label{g1ksp2}
g_{1, \BZ_k} (t, x_1, x_2) 
= \sum_{n_1, n_2 =0}^\infty \sum_{a=0}^{p(n_1,n_2)} \sum_{r=0}^{k-1}~[k(n_1+n_2) +2(r -a), a] ~ t^{k(n_1+n_2) +2r}~,\nn \\
\eea
where $p(n_1,n_2) = \mathrm{min}\{ kn_1+r, kn_2+r \}$, and $x_1,~x_2$ are the $Sp(2)$ fugacities.  

\subsection{$\hat{D}_{k+2}$ Orbifolds}
Let us consider the group $\hat{D}_{k+2}$ which is a subgroup of $SO(8)$.  It is generated by
\begin{equation} \label{gend}
\left( \begin{array}{cc} w I_{4\times 4} & 0\\0 & w^{-1} I_{4\times 4} \end{array} \right)   \qquad  , \qquad 
\left( \begin{array}{cc} 0 &  i J_{4\times 4} \\ -i  J_{4\times 4}& 0\end{array} \right)
\end{equation}
where $w^{2k}=1$ and $J_{4\times 4}$ is the four by four symplectic matrix\footnote{These generators are consistent with the generators taken from Equations (3.9) and (3.10) of \cite{BFHH} by taking a 2 by 2 block matrix and composing it with the 4 by 4 matrix that has an identity in the diagonal components, $J$ in the upper block and $-J$ in the lower block. A similar construction follows for the other non-abelian subgroups of $SU(2)$ as stated explicitly below.}.
The matrices in the previous formula are acting on the vector representation
$[1,0,0,0]$ of $Spin(8)$ in a complex notation where it decomposes as a  fundamental $[1,0,0]$ plus anti-fundamental $[0,0,1]$ representation of $SU(4)$. The global symmetry group $SU(4)\times U(1)_B$ is reduced by the projection 
to $Sp(2)$, which is simply the group of $SU(4)$ matrices satisfying the condition $J g =g^* J$.
%The
%generator of the abelian subgroup $Z_{2k}$ acts as before on the baryonic fugacity $b\rightarrow w  b%$ while the other dihedral generator acts as
%\begin{equation}
%b\rightarrow \frac{1}{b}\,\qquad z_1\rightarrow z_3\, , \,\, z_2\rightarrow z_2\, , \,\, z_3\rightarrow z_1 
%\end{equation}

\paragraph{General partition function for $\hat{D}_{k+2}$.}  It can be shown \cite{Ikeda} that substituting \eref{gend} into \eref{dismo} gives the partition function for $N=1$ and $\Gamma= \hat{D}_{k+2}$ for an arbitrary $k$. This substitution and the substitution for the other non-abelian groups is consistent with the formulas in Table (3.9) of \cite{BFHH}.
\beq \label{1d2k}
g_{1, \hat{D}_{k+2}}(t,x_1,x_2) = \frac{1}{2} g_{1, \BZ_{2k}}(t,x_1,x_2) + g_{1, \BZ_{4}}(t,x_1,x_2) - \frac{1}{2} g_{1, \BZ_{2}}(t,x_1,x_2)~.
\eeq
The rationale for this formula is that we can consider $\hat{D}_{k+2}$ as composed of a subgroup $\BZ_{2k}$ and $k$ subgroups of $\BZ_4$, each with common intersection $\BZ_2$. (\ref{1d2k}) is a surgery formula, as in \cite{Hanany:2006uc}, for this decomposition.

\paragraph{An example of $\hat{D}_4$.} Substituting $k=2$ into \eref{1d2k} and using \eref{g12}, \eref{g14}, we find that the unrefined partition function is given by 
\bea 
g_{1,\hat{D}_4}(t, 1, 1) &=& \frac{1+2 t^2+68 t^4 +78 t^6 + 214 t^8 +78 t^{10} +68 t^{12} +2 t^{14} +t^{16}}{(1-t^2)^7(1+t^2)^4} \nn\\
&=& 1+5 t^2+84 t^4+324 t^6+1221 t^8+3185 t^{10}+\ldots~.
\eea
The plethystic logarithm is given by
\beq 
g_{1,\hat{D}_4}(t, 1, 1) = 5 t^2+69 t^4-56 t^6-2019 t^8+3368 t^{10}+\ldots~.
\eeq
The coefficient 5 of $t^2$ indicates that there are 5 generators transforming in the $[0,1]$ representation, and the coefficient 69 of $t^4$ is the dimension of the representation $[4,0] + [2,1] -[0,0]$.  We note that the correction [0,0], which is contained in the decomposition of $\mathrm{Sym}^2 [0,1]$, simply indicates that there is a relation of order $t^4$.  Thus, the generators of this theory transform in the representations [0,1], [2,1] and [4,0].

\paragraph{General formulae.} Substituting \eref{unrefzk} into \eref{1d2k}, we obtain the general unrefined partition function for $\hat{D}_{k+2}$:
\bea
g_{1,\hat{D}_{k+2}} (t,1,1) &=& \frac{1}{3 \left(1-t^2\right)^6 \left(1+t^2\right)^4 \left(1-t^{2 k}\right)^4} \times \nn \\
&& \Big[3+9 t^2+108 t^4+90 t^6+195 t^8+45 t^{10}+30 t^{12}+\left(-9+11 k+12 k^2+4 k^3\right) t^{2 k} \nn \\
&& +\left(9-22 k+16 k^3\right) t^{4 k}+\left(-3+11 k-12 k^2+4 k^3\right) t^{6 k}-\big(3+71 k+36 k^2 \nn \\
&& +4 k^3 \big) t^{6 (2+k)}+ \big(63+142 k-16 k^3\big) t^{4 (3+k)}+3 \big(-9-49 k+4 k^2+4 k^3\big) t^{2 (5+k)} \nn \\
&& -\left(81+71 k-36 k^2+4 k^3\right) t^{2 (6+k)}+\left(3-11 k+12 k^2-4 k^3\right) t^{2 (7+k)}\nn \\
&& +\left(3+71 k+36 k^2+4 k^3\right) t^{2+2 k}-3 \left(93-49 k-4 k^2+4 k^3\right) t^{4+2 k} \nn \\
&& -3 \left(25-29 k+20 k^2+4 k^3\right) t^{6+2 k}+3 \left(-165-29 k-20 k^2+4 k^3\right) t^{8+2 k} \nn \\
&& +\left(-63-142 k+16 k^3\right) t^{2+4 k}-3 \left(-63+98 k+16 k^3\right) t^{4+4 k} \nn \\
&& -3 \left(105+58 k+16 k^3\right) t^{6+4 k} +3 \left(105+58 k+16 k^3\right) t^{8+4 k}\nn \\
&& +3 \left(-63+98 k+16 k^3\right) t^{10+4 k}+\left(-9+22 k-16 k^3\right) t^{14+4 k} \nn \\
&& +\left(81+71 k-36 k^2+4 k^3\right) t^{2+6 k}-3 \left(-9-49 k+4 k^2+4 k^3\right) t^{4+6 k} \nn \\
&& +\left(495+87 k+60 k^2-12 k^3\right) t^{6+6 k}+3 \left(25-29 k+20 k^2+4 k^3\right) t^{8+6 k} \nn\\
&& +3 \left(93-49 k-4 k^2+4 k^3\right) t^{10+6 k}-\left(-9+11 k+12 k^2+4 k^3\right) t^{14+6 k} \nn \\
&& -30 t^{2+8 k}-45 t^{4+8 k}-195 t^{6+8 k}-90 t^{8+8 k}-108 t^{10+8 k}-9 t^{12+8 k} -3 t^{14+8 k} \Big]~,\qquad
\eea
where we note that this formula is consistent with the above specific examples.
An explicit expression for the refined partition function is given by
\beq 
g_{1,\hat{D}_4}(t, x_1, x_2) = \sum_{n=0}^\infty\sum_{p=0}^\infty\sum_{j=0,\ne n-1}^n [2 n + 4 p -2 j,j] t^{2 n + 4 p} .
\eeq

\subsection{$\hat{E}_{6}$ Orbifold}
Let us consider the group $\hat{E}_{6}$ which is a subgroup of $Spin(8)$.  
It is generated by
\begin{equation} \label{gene6}
\frac{1}{2} \left ( \begin{array}{cc} (-1+i) I_{4\times 4} & (-1+i) J_{4\times 4}\\ - (1+i) J_{4\times 4} & (-1-i) I_{4\times 4} \end{array} \right )   \qquad  , \qquad 
\left ( \begin{array}{cc}  i I_{4\times 4} & 0 \\ 0 & -i I_{4\times 4}\end{array} \right )
\end{equation}

\paragraph{Full partition function for $\hat{E}_{6}$.}  It can be shown \cite{Ikeda} that the partition function for $N=1$ and $\Gamma= \hat{E}_{6}$ is
\beq \label{1e6}
g_{1, \hat{E}_{6}}(t,x_1,x_2) = g_{1,\BZ_6}(t,x_1,x_2)+\frac{1}{2} \left( g_{1,\BZ_4}(t,x_1,x_2) - g_{1,\BZ_2}(t,x_1,x_2) \right)~.
\eeq
Substituting \eref{g1ksp2} into \eref{1e6}, we obtain the full partition function for $\hat{E}_{6}$.

\paragraph{The unrefined partition function.}  We obtain a simpler expression if the  $x$'s are set to unity:
\bea
g_{1, \hat{E}_{6}}(t,1,1) &=& \frac{1}{\left(1-t^2\right)^7 \left(1+2 t^2+2 t^4+t^6\right)^4} \times \nn\\ 
&& (1+6 t^2+16 t^4+106 t^6+487 t^8+996 t^{10}+1532 t^{12}+2332 t^{14}+2872 t^{16}+ \nn \\
&& 2332 t^{18}+1532 t^{20}+996 t^{22}+487 t^{24}+106 t^{26}+16 t^{28}+6 t^{30}+t^{32}) \nn \\
&=& 1+5 t^2+14 t^4+114 t^6+451 t^8+975 t^{10}+\ldots~.
\eea
The plethystic logarithm is given by
\beq 
g_{1,\hat{E}_6}(t, 1, 1) =5 t^2-t^4+84 t^6-24 t^8-172 t^{10}+\ldots~.
\eeq
The coefficient 5 of $t^2$ indicates that there are 5 generators transforming in the $[0,1]$ representation,  the coefficient $-1$ of $t^4$ indicates that there is a relation transforming in the trivial representation, and the coefficient 84 of $t^4$ indicates that there are 84 generators transforming in the $[6,0]$ representation. Thus, the first generators of this theory transform in the representations $[0,1]$ and $[6,0]$.

\subsection{$\hat{E}_{7}$ Orbifold}
Let us consider the group $\hat{E}_{7}$ which is a subgroup of $Spin(8)$. 
It is generated by
\begin{equation} \label{gene7}
\frac{1}{2} \left ( \begin{array}{cc} (-1+i) I_{4\times 4} & (-1+i) J_{4\times 4}\\ - (1+i) J_{4\times 4} & (-1-i) I_{4\times 4} \end{array} \right )   \qquad  , \qquad 
\frac{1}{\sqrt{2}}\left ( \begin{array}{cc}  ( 1 + i) I_{4\times 4} & 0 \\ 0 & (1 - i)  I_{4\times 4} \end{array} \right )
\end{equation}
%\begin{equation} \label{gene7}
%\frac{1}{2} \left ( \begin{array}{cc} (-1+i) I_{4\times 4} & (-1+i) I_{4\times 4}\\(1+i) I_{4\times 4} & (-1-i) I_{4\times 4} \end{array} \right )   \qquad  , \qquad 
%\left ( \begin{array}{cc} 0 &  i J_{4\times 4} \\ -i  J_{4\times 4}& 0\end{array} \right )
%\end{equation}
\paragraph{Full partition function for $\hat{E}_{7}$.}  It can be shown \cite{Ikeda} that the partition function for $N=1$ and $\Gamma= \hat{E}_{7}$ is
\beq \label{1e7}
g_{1, \hat{E}_{7}}(t,x_1,x_2) =  \frac{1}{2} (g_{1,\BZ_8}(t,x_1,x_2) + g_{1,\BZ_6}(t,x_1,x_2)  + g_{1,\BZ_4}(t,x_1,x_2)  - g_{1,\BZ_2}(t,x_1,x_2) )~.
\eeq
Substituting \eref{g1ksp2} into \eref{1e7}, we obtain the full partition function for $\hat{E}_{7}$.

\paragraph{The unrefined partition function.}  We obtain a simpler expression if the  $x$'s are set to unity:
\bea
g_{1, \hat{E}_{7}}(t,1,1) &=& \frac{1}{\left(1-t^2\right)^7 \left(1+2 t^2+3 t^4+3 t^6+2 t^8+t^{10}\right)^4 }  \times \nn \\ 
&& (1+6 t^2+20 t^4+46 t^6+242 t^8+686 t^{10}+1921 t^{12}+3602 t^{14}+6037 t^{16}+ \nn \\ 
&& 8672 t^{18}+11947 t^{20}+14252 t^{22}+15728 t^{24}+14252 t^{26}+11947 t^{28}+ \nn \\
&& 8672 t^{30}+6037 t^{32}+3602 t^{34}+1921 t^{36}+686 t^{38}+242 t^{40}+46 t^{42}+\nn \\
&& 20 t^{44}+6 t^{46}+t^{48} ) \nn \\
&=& 1+5 t^2+14 t^4+30 t^6+220 t^8+520 t^{10}+\ldots~.
\eea
The plethystic logarithm is given by
\beq 
g_{1,\hat{E}_7}(t, 1, 1) =5 t^2-t^4+165 t^8-396 t^{10}+\ldots~.
\eeq
The coefficient 5 of $t^2$ indicates that there are 5 generators transforming in the $[0,1]$ representation,  the coefficient $-1$ of $t^4$ indicates that there is a relation transforming in the trivial representation, and the coefficient 165 of $t^8$ indicates that there are 84 generators transforming in the $[8,0]$ representation. Thus, the first generators of this theory transform in the representations $[0,1]$ and $[8,0]$.

\subsection{$\hat{E}_{8}$ Orbifold}
Let us consider the group $\hat{E}_{8}$ which is a subgroup of $Spin(8)$.  It is generated by the same generators as $\hat{E}_{6}$ with the addition of
\begin{equation} \label{gene8}
 \frac{1}{4}\left ( \begin{array}{cc} 2 i  I_{4\times 4} &    ((1-\sqrt{5})-i(1+\sqrt{5}))J_{4\times 4}\\    ((1-\sqrt{5})+i(1+\sqrt{5}))J_{4\times 4} & -2 i I_{4\times 4}\end{array} \right )
\end{equation}

\paragraph{Full partition function for $\hat{E}_{8}$.}  It can be shown \cite{Ikeda} that the partition function for $N=1$ and $\Gamma= \hat{E}_{8}$ is
\beq \label{1e8}
g_{1, \hat{E}_{8}}(t,x_1,x_2) =  \frac{1}{2} (g_{1,\BZ_{10}}(t,x_1,x_2) + g_{1,\BZ_6}(t,x_1,x_2) + g_{1,\BZ_4}(t,x_1,x_2)- g_{1,\BZ_2}(t,x_1,x_2))~.
\eeq
Substituting \eref{g1ksp2} into \eref{1e8}, we obtain the full partition function for $\hat{E}_{8}$.

\paragraph{The unrefined partition function.}  We obtain a simpler expression if the $x$'s are set to unity:
\bea
g_{1, \hat{E}_{8}}(t,1,1) &=& \frac{1}{\left(1-t^2\right)^7 \left(1+3 t^2+5 t^4+6 t^6+6 t^8+5 t^{10}+3 t^{12}+t^{14}\right)^4}\times \nn \\
&& (1+10 t^2+50 t^4+166 t^6+410 t^8+798 t^{10}+1711 t^{12}+4970 t^{14}+ \nn \\
&& 14024 t^{16}+30920 t^{18}+53137 t^{20}+75728 t^{22}+97846 t^{24}+124794 t^{26}+\nn\\
&& 160086 t^{28}+194598 t^{30}+209502 t^{32}+194598 t^{34}+160086 t^{36}+124794 t^{38}+\nn \\
&& 97846 t^{40}+75728 t^{42}+53137 t^{44}+30920 t^{46}+14024 t^{48}+4970 t^{50}+1711 t^{52}+\nn \\
&& 798 t^{54}+410 t^{56}+166 t^{58}+50 t^{60}+10 t^{62}+t^{64}) \nn\\
&=& 1+5 t^2+14 t^4+30 t^6+55 t^8+91 t^{10}+\ldots~.
\eea
The plethystic logarithm is given by
\beq 
g_{1,\hat{E}_8}(t, 1, 1) =5 t^2-t^4+455 t^{12}-1170 t^{14}+\ldots~.
\eeq
The coefficient 5 of $t^2$ indicates that there are 5 generators transforming in the $[0,1]$ representation,  the coefficient $-1$ of $t^4$ indicates that there is a relation transforming in the trivial representation, and the coefficient 455 of $t^{12}$ indicates that there are 455 generators transforming in the $[12,0]$ representation. Thus, the first generators of this theory transform in the representations $[0,1]$ and $[12,0]$.

%The full partition function for $\mathbb{R}/\Gamma$ depends on the parameter $t$ 
%counting dimension and on a set of $Sp(2)$ weights $x_1,x_2$. 
%It is given by the following discrete Molien formula
%\begin{equation}
%g_{1,D_{k+2}} (t, x_1, x_2) = \frac{1}{4k} \sum_{\gamma\in D_{k+2}} g_{1,\BZ_1} (t, \gamma(\{ b, z_1, z_2, %z_3\}))\Big |_{b=1,z_1=x_1,z_2=x_2^2,z_3=x_1} 
%\end{equation}
%obtained by averaging over all the elements of the discrete group. 
%After the projection, we restrict the result to $Sp(2)$ weights;
%we also set $b=1$ since the baryonic fugacity is no more a good quantum 
%number.

%Recalling \eref{11su4} and performing the projection we can write
%\begin{eqnarray}
%g_{1,D_{k+2}} (t,x_1,x_2) &=& \sum_{m-n=0 (mod k)} t^{2n+kp}\frac{[n,0,m]+[m,0,n]}{2} \nonumber\\
%&&\sum_{n=0}^\infty t^{2n} \sum_{j=0}^n [2n-2j,j] + 2\sum_{n=0,p=0}^\infty t^{2n+kp} \sum_{j=0}^n [2n%+kp-2j,j]
%\end{eqnarray}
%where now $[n,m]$ denotes a representation of $Sp(2)$. {\bf need to recheck}
%Generators etc...

%In terms of the parameter $t$ we have, for example, for $D_4$
%\begin{equation} 
%g_{1,D_4}(t) = \frac{1+12 t^2+108 t^4 +212 t^6 + 358 t^8 +212 t^{10} +108 t^{12} +t^{16}}{(1-%t^2)^7(1+t^2)^4}
%\end{equation}

%%%%%%%%%%%%%%%%%
\section{Higher $N$ Theories} \setall

Having dealt with various $N=1$ theories, we turn to the problem of counting operators in higher $N$ case.  

\paragraph{The moduli space.}  We will denote the moduli space for $N$ branes on $\BR^8$ by $S^N(\BR^8)$. Restricting to $S^7$, we quotient this out by the non-compact direction $\BR^+$ to get ${\cal M}_N (S^7) = S^N(\BR^8)/\BR^+$. This moduli space has a real dimension $8N-1$. 

\paragraph{The grand canonical partition function.}  We use the plethystic exponential and write down the generating function for higher values of $N$ by introducing a fugacity $\nu$ for the number of M2-branes. We may also choose to count operators on $S^7$. For this purpose we first write the grand canonical partition function for $\BR^8$,
\begin{equation}
g(\nu; t, y_1,y_2,y_3, y_4;  \BR^8 ) = \PE \left [ \nu g_{1,\Gamma} (t, y_1,y_2,y_3, y_4; \BR^8 ) \right ],
\end{equation}
that has an expansion in terms of partition functions for a fixed number of branes
\begin{equation}
g(\nu; t, y_1,y_2,y_3, y_4;  \BR^8 ) = \sum_{N=0}^\infty \nu^N g_{N,\Gamma} \left (t, y_1,y_2,y_3, y_4; S^N(\BR^8) \right ) 
\end{equation}
Explicitly, the formulae for the first few $N$ are as follows:
\bea \label{explicitN}
g_{2, \Gamma} (t, y; S^2(\BR^8)) &=&  \frac{1}{2} \left[ g_{1,\Gamma}(t, y; \BR^8)^2 + g_{1,\Gamma}(t^2, y^2; \BR^8) \right]~, \nn \\
g_{3, \Gamma} (t, y; S^3(\BR^8)) &=& \frac{1}{6} \left[ g_{1,\Gamma}(t, y; \BR^8)^3 + 3 g_{1,\Gamma}(t, y) g_{1,\BZ_k}(t^2, y^2; \BR^8) + 2 g_{1,\Gamma}(t^3, y^3; \BR^8) \right]~, \nn\\
\eea
where we have written $y_1,~y_2,~y_3,~y_4$ collectively as $y$.

\paragraph{Operators on $S^7$.} The projection to protected operators is more complicated than for the $N=1$ case since products of short multiplets are not necessarily short. We just remove an overall trace and regard these partition functions
as counting real functions on the moduli space for $N$ branes. 
%We leave for future work the investigation of their supersymmetric properties.
{One needs to note that {\bf the restriction to a fixed radius should be done only once and should not be symmetrised over.}
We get the reduced grand canonical partition function,
\begin{equation} \label{s7}
g_{\Gamma} (\nu; t, y_1,y_2,y_3, y_4; S^7) = (1-t^2) \PE \left [ \nu g_{1,\Gamma} (t, y_1,y_2,y_3, y_4; \BR^8 ) \right ].
\end{equation}

\paragraph{An example of $N=2$ and $k=1$.} Using the first formula in \eref{explicitN} together with \eref{s7}, we find that the coefficient of $\nu^2$ in the above expression is 
\beq \ba{rcl}
g_{2, \BZ_1} (t, y; \CM_2(S^7)) %&=& \frac{1}{2} (1-t^2) \left[ g_{1,\BZ_1}\left(t,y;\BR^8\right)^2 + g_{1,\BZ_1}\left(t^2,y^2;\BR^8\right) \right] \nn \\
&=& \frac{1-t^2}{2\left(1-\frac{t}{y_1}\right){}^2 \left(1-t y_1\right){}^2 \left(1-\frac{t y_1}{y_2}\right){}^2 \left(1-\frac{t y_2}{y_1}\right){}^2 \left(1-\frac{t y_2}{y_3 y_4}\right){}^2 \left(1-\frac{t y_3}{y_4}\right){}^2 \left(1-\frac{t y_4}{y_3}\right){}^2 \left(1-\frac{t y_3 y_4}{y_2}\right){}^2} \nn\\
&& +\frac{1-t^2}{2\left(1-\frac{t^2}{y_1^2}\right) \left(1-t^2 y_1^2\right) \left(1-\frac{t^2 y_1^2}{y_2^2}\right) \left(1-\frac{t^2 y_2^2}{y_1^2}\right) \left(1-\frac{t^2 y_2^2}{y_3^2 y_4^2}\right) \left(1-\frac{t^2 y_3^2}{y_4^2}\right) \left(1-\frac{t^2 y_4^2}{y_3^2}\right) \left(1-\frac{t^2 y_3^2 y_4^2}{y_2^2}\right)}~.
%g_{3, 1} (t, y) &=& \frac{1}{6} (1-t_1 t_2) \left[ g_{1,\BZ_1}(t,y)^3 + 3 g_{1,\BZ_1}(t,y) g_{1,\BZ_1}(t^2,y^2) + 2 g_{1,\BZ_1}(t^3,y^3) \right]~, \nn \\
%g_{4,1} (t, y) &=& \frac{1}{24}(1-t_1 t_2) [ g_{1,\BZ_1}(t, y)^4 + 6 g_{1,\BZ_1}(t,y)^2 g_{1,\BZ_1}(t^2,y^2) + 3 g_{1,\BZ_1}(t^2, y^2)^2 + \nn \\ 
%&& 8 g_{1,\BZ_1}(t,y) g_{1,\BZ_1}(t^3,y^3) + 6 g_{1,\BZ_1}(t^4,y^4) ]~.
\ea
\eeq
%where $y$ collectively denotes $y_1, y_2, y_3, y_4$.
This can be expanded in terms of irreducible representations of $Spin(8)$ as
\beq
g_{2, \BZ_1} (t, y; \CM_2(S^7))  = 1 + \left( 2[2,0,0,0]+1\right) t^2 + \left( 2[3,0,0,0] + 2[1,0,0,0] +[1,1,0,0] \right)t^3 + \ldots~.
\eeq
Note that at order 2 we again find the singlet operator of the form $\mathrm{Tr}(\phi_i \phi_i)$ or $\mathrm{Tr}(\phi_i)\mathrm{Tr}(\phi_i)$, either of which is unprotected. This is the simplest example which demonstrates that higher $N$ generating functions do not count protected operators.
%Using the decomposition \eref{decomp}, we see that this theory has a simple expansion in terms of characters of $SU(4)$:
%\bea
%g_{2,\BZ_1} (t,b, z_1,z_2,z_3; \widetilde{S^2(\BR^8)}) = \sum_{n=0}^\infty \sum_{q=0}^{n} \sum_{r=0}^{2n-2q} b^{2n-2q-2r} [2n-2q-r,0,r] t^{2n}~.
%\eea
%\bea
%g_{2,\BZ_1} (t,b, z_1,z_2,z_3; \widetilde{S^2(\BR^8)}) = \sum_{n=0}^\infty \sum_{q=0}^{n} \sum_{r=0}^{2n-2q} b^{2n-2q-2r} [2n-2q-r,0,r] t^{2n}~.
%\eea
When setting the $Spin(8)$ chemical potentials to zero we find
\bea
g_{2,\BZ_1} (t, 1,1,1,1; \CM_2(S^7))) &=& \frac{1+28 t^2+70 t^4+28 t^6+t^8}{(1-t)^{15} (1+t)^7} \nn\\
&=& 1+8 t+71 t^2+400 t^3+1884 t^4+7344 t^5+\ldots
\eea
with the pole of order 15 at $t=1$ indicating that the reduced moduli space is indeed $8\times2-1=15$ real dimensional.
The plethystic logarithm of this expression is 
\bea
\PL[g_{2,\BZ_1} (t, 1,1,1,1; \CM_2(S^7)))] = 8 t+35 t^2-336 t^4+5376 t^6-101856 t^8+\ldots~.
\eea
This indicates that there are 8 generators transforming in the $Spin(8)$ representation $[1,0,0,0]$ at order $t$, 35 generators transforming in $[2,0,0,0]$ at order $t^2$, and 336 relations transforming in $[0,2,0,0]+[2,0,0,0]+[0,0,0,0]$ at order $t^6$.

\paragraph{An example of $N=2$ and $k=2$.}  For simplicity, let us work with unrefined partitions. Starting from the case of $N=1$, we have
\bea
g_{1,\BZ_2}(t, 1, 1, 1, 1; \BR^8) = \frac{g_{1,\BZ_2}(t, 1, 1, 1, 1; S^7)}{1-t^2}  =  \frac{1+28 t^2+70 t^4+28 t^6+t^8}{\left(1-t^2\right)^8}~,
\eea
where we have used \eref{g12} in the second equality.  Using the first formula in \eref{explicitN} and restricting to $S^7$, we find that
\bea
g_{2,\BZ_2}(t, 1, 1, 1, 1;  \CM_2(S^7)) &=& \frac{1}{\left(1-t^2\right)^{15} \left(1+t^2\right)^8} \times (1+28 t^2+728 t^4+6356 t^6+34140 t^8+ \nn \\
&& 110300 t^{10}+254184 t^{12}+403508 t^{14}+478662 t^{16}+403508 t^{18}+\nn \\
&& 254184 t^{20}+110300 t^{22} + 34140 t^{24}+6356 t^{26}+728 t^{28}+28 t^{30}+t^{32}) \nn \\
&=& 1+35 t^2+960 t^4+12600 t^6+109230 t^8+\ldots~.
\eea
The plethystic logarithm of this expression is
\bea
\PL[g_{2,\BZ_2} (t, 1,1,1,1; \CM_2(S^7)))] = 35 t^2+330 t^4-6720 t^6+8100t^8 + \ldots~.
\eea
This indicates that there are 35 generators transforming in the $Spin(8)$ representation $[2,0,0,0]$ at order $t^2$, and 330 generators transforming in $[4,0,0,0] + [2,0,0,0] + [0,0,0,0]$ at order $t^4$.
%Factoring out the centre of mass motion using \eref{FCM3}, we obtain
%\bea
%g_{2,\BZ_2}(t, 1, 1, 1, 1; \widetilde{S^2(\BR^8)}) &=& \frac{1}{\left(1-t^4\right)^8  \left(1+28 t^2+70 t^4+28 t^6+t^8\right)} \times (1+28 t^2+728 t^4+ \nn \\
%&& 6356 t^6+34140 t^8+110300 t^{10}+254184 t^{12}+403508 t^{14}+478662 t^{16}+ \nn \\ 
%&& 403508 t^{18}+254184 t^{20}+110300 t^{22}+34140 t^{24}+6356 t^{26}+728 t^{28}+\nn \\ 
%&& 28 t^{30}+t^{32})~.
%\eea

%However, if we set $b$ and $z$'s to unity, then
%\bea
%g_{2,\BZ_2} (t, 1, 1, 1, 1) &=& \frac{1}{\left(1-t^2\right)^{14} \left(1+t^2\right)^7} (1+28 t^2+693 t^4+5712 t^6+28197 t^8+82796 t^{10}+ \nn\\ 
%& & 169897 t^{12}+236096 t^{14}+239331 t^{16}+167412 t^{18}+84287 t^{20}+27504 t^{22}+ \nn \\
%& & 5943 t^{24}+644 t^{26}+35 t^{28} )~.
%\eea
%The plethystic logarithm of this expression is
%\beq
%\PL[g_{2,\BZ_2} (t, 1, 1, 1, 1)] = 35 t^2 + 294 t^4 - 6384 t^6 + 17634 t^8 + \ldots~.
%\eeq
%This indicates that there are 35 generators transforming in the $SO(8)$ representation $[2,0,0,0]$ at order $t^2$, 294 generators transforming in the representation $[4,0,0,0]$ at order $t^4$, and 6384 relations transforming in the representation $[4, 0, 0, 0] + [2, 2, 0, 0] + [2, 1, 0, 0] + 2 [2, 0, 0, 0] + [0, 2, 0, 0] + [0, 0, 2, 2] + [0, 0, 0, 0]$ at order $t^6$.

\paragraph{The palindromic property.} Note that the previous partition functions are palindromic. This 
happened for all the partition functions for one or more membranes that we encountered in this paper. It can be explained as follows. Recall that the palindromic property characterizes Calabi-Yau (Gorenstein) singularities \cite{Forcella:2008bb}. Although we are considering real coordinates, the partition function for a membrane on $\mathbb{R}^8/\Gamma$ can be equivalently considered as a partition function for holomorphic functions on the complexification $\mathbb{C}^8/\Gamma$
which is indeed a non-compact Calabi-Yau singularity. Analogously, for $N$ membranes, we deal with the symmetric product of $N$ Calabi-Yau four-folds which is also a non-compact Calabi-Yau singularity\footnote{This should be contrasted with the three dimensional case where moduli spaces for $N>1$ are  
not Calabi-Yau \cite{Forcella:2008bb} since symmetrized products of odd dimensional Calabi-Yaus are not.}.

%\subsection{$\BZ_k$ Orbifold Actions on Higher $N$ Theories}
%The partition function for higher values of $N$ on $\BR^8$ is given by the expansion
%\begin{equation}
%\sum_{N=0}^\infty \nu^N g_{N,\BZ_k} (t, b, z_1, z_2, z_3; \BR^8) = \PE [ \nu g_{1,\BZ_k}(t, b, z_1, z_2, z_3; \BR^8) ] 
%\end{equation}
%Explicitly, the formulae for the first few $N$ are as follows:
%\bea \label{explicitN}
%g_{2, k} (t, b, z; \BR^8) &=&  \frac{1}{2} \left[ g_{1,\BZ_k}(t, b, z; \BR^8)^2 + g_{1,\BZ_k}(t^2, b^2, z^2; \BR^8) \right]~, \nn \\
%g_{3, k} (t, b, z; \BR^8) &=& \frac{1}{6} \left[ g_{1,\BZ_k}(t,b,z; \BR^8)^3 + 3 g_{1,\BZ_k}(t,b,z) g_{1,\BZ_k}(t^2,b^2,z^2; \BR^8) + 2 g_{1,\BZ_k}(t^3,b^3,z^3; \BR^8) \right]~, \nn\\
%\eea
%where we have written $z_1,~z_2,~z_3$ collectively as $z$.

%The restriction onto $S^7$ can be done by simply using the formula:
%\beq \label{ress7}
%g_{N,\BZ_k} (t, b, z; S^7) = (1-t^2) g_{N,\BZ_k} (t, b, z; \BR^8)~.
%\eeq
%As a check, the order of the pole at $t=1$ should be equal to the dimension of moduli space $\CM_N(S^7)$, which is $8N-1$.

%We also remind the reader that the centre of mass motion can be factored out by using formulae \eref{FCM1} or \eref{FCM2}:
%\begin{equation} \label{FCM3}
%g_{N,\BZ_k} (t, b, z; \widetilde{S^N(\BR^8)}) = \frac{g_{N,\BZ_k} (t, b, z;\BR^8)}{g_{1,\BZ_k} (t, b,z ;\BR^8)} =\frac{g_{N,\BZ_k} (t, b, z;S^7)}{g_{1,\BZ_k} (t, b,z ;S^7)}~.
%\end{equation} 
%As a check, the order of the pole at $t=1$ should be equal to the dimension of moduli space $\CM_N(\widetilde{S^N(\BR^8)})$, which is $8(N-1)$.

\section*{Acknowledgements}
A.~H.~ and A.~Z.~ thank the Galileo Galilei Institute for
Theoretical Physics for the hospitality and the INFN for partial support
during the completion of this work.
A.~Z.~ is supported
in part by INFN and MIUR under contract 2005-024045-004 and
2005-023102 and by the European Community's Human Potential Program
MRTN-CT-2004-005104.
N.~M. would like to express his gratitude to the following:
his family for the warm encouragement and support; his colleagues Alexander Shannon, Benjamin Withers, William Rubens, and Sam Kitchen
for valuable discussions;  
and, finally, the DPST Project and the Royal Thai Government for funding his research.

\appendix

\section{Character Computations: Plethystic Amusement} \setall
In this section we present an efficient method for computing all characters of a given group. This method is good when the rank of the group is small enough. We will demonstrate this method for the group $SO(5)$, with $Sp(2)$ characters being obtained as by-products.  Moreover, $SU(4)$ characters will be mentioned at the end of this section.  This method can be repeated for the group $SO(8)$.

\paragraph{$SO(5)$ characters.} Let us start by looking at a generic representation of $SO(5)$ of the form $[n_1,n_2]_{SO(5)}$ with dimension formula \cite{FH} given by
\begin{equation} \label{dimso5}
\dim [n_1,n_2]_{SO(5)} = \frac{(n_1+1)(n_2+1)(n_1+n_2+2)(2n_1+n_2+3)}{6} .
\end{equation}
We can introduce a formal generating function
\begin{equation} \label{Acharexp}
g_{SO(5)} (t_1, t_2; w_1, w_2) = \sum_{n_1=0}^\infty \sum_{n_2=0}^\infty [n_1,n_2]_{SO(5)} t_1^{n_1} t_2^{n_2}~,
\end{equation}
where $t_1, t_2$ are weights which keep track with the highest weight representation of the group, while $w_1, w_2$ are the $SO(5)$ fugacities.
The function $g_{SO(5)}$ can be summed easily. One can write down the expression when the $SO(5)$ chemical potentials are set to zero,
\begin{equation}\label{APE1}
g_{SO(5)} (t_1, t_2; 1, 1) =\frac{ 1-t_1^2 - 4 t_1 t_2 +t_1t_2^2  +4 t_1^2 t_2  -t_2^2 t_1^3}{(1-t_1)^5(1-t_2)^4},
\end{equation}
and immediately realize that this can be written in terms of the simple representations of $SO(5)$,
\begin{equation} \label{APE2}
g_{SO(5)} (t_1, t_2; w_1, w_2) ={( 1-t_1^2 - [0,1] t_1 t_2 +t_1t_2^2  + [0,1] t_1^2 t_2 - t_1^3 t_2^2 )} \PE \left [ [1,0]  t_1 + [0,1] t_2 \right ]~,
\end{equation}
One can compute the character of the representation with highest weights $[n_1, n_2]$ as the coefficient of the $t_1^{n_1} t_2^{n_2}$ term in the power expansion of $g_{SO(5)}$. For completeness we record here a possible explicit form for the two representations which are needed to evaluate this expansion,
\begin{equation} \label{1001}
[1,0]_{SO(5)} = \frac{w_2^2}{w_1}+w_1+\frac{1}{w_1}+1+\frac{w_1}{w_2^2}~, \quad [0,1]_{SO(5)} = \frac{w_1}{w_2}+w_2+\frac{1}{w_2}+\frac{w_2}{w_1}~.
\end{equation}
As an example, let us consider the $SO(5)$ representation $[1,1]$.  Equation \eref{Acharexp} suggests that the character of $[1,1]$ is simply the coefficient of $t_1 t_2$ in the power expansion of the right hand side of \eref{APE2}.  Substituting \eref{1001} into \eref{APE2} and expanding it as a power series, we find that the coefficient of $t_1 t_2$ is
\beq
[1,1]_{SO(5)} =  \frac{w_1}{w_2^3}+\frac{w_1^2}{w_2^3}+\frac{2}{w_2}+\frac{1}{w_1 w_2}+\frac{2 w_1}{w_2}+\frac{w_1^2}{w_2}+2 w_2+\frac{w_2}{w_1^2}+\frac{2 w_2}{w_1}+w_1 w_2+\frac{w_2^3}{w_1^2}+\frac{w_2^3}{w_1}~.
\eeq

\paragraph{$Sp(2)$ characters: By-products.}   The dimension formula \cite{FH} of the representation $[n_1,n_2]_{Sp(2)}$ is given by
\begin{equation}
\dim [n_1,n_2]_{Sp(2)} = \frac{(n_1+1)(n_2+1)(n_1+n_2+2)(n_1+2n_2+3)}{6}~.
\end{equation}
Observe that this is simply the interchange of $n_1$ and $n_2$ in formula \eref{dimso5}, \emph{i.e.}
\beq
 \dim [n_1,n_2]_{Sp(2)} = \dim [n_2,n_1]_{SO(5)}~.
\eeq
Therefore, the corresponding \eref{APE1} for $Sp(2)$ is
\beq
 g_{Sp(2)} (t_1, t_2; 1, 1) = g_{SO(5)} (t_2, t_1; 1, 1)~.
\eeq
We note that $\dim[1,0]_{Sp(2)} = 4$ and $\dim[0,1]_{Sp(2)} = 5$.
According to \eref{APE2}, we have
\bea 
g_{Sp(2)} (t_1, t_2; w_1, w_2) &=& {( 1-t_2^2 - [1,0] t_2 t_1 +t_2t_1^2  + [1,0] t_2^2 t_1 - t_2^3 t_1^2 )} \PE \left [ [1,0]  t_2 + [0,1] t_1 \right ] \nn \\
&=&g_{SO(5)} (t_2, t_1; w_2, w_1)~.
\eea
Thus, we arrive at an amusing relation between irreducible representations of $Sp(2)$ and $SO(5)$:
\beq
[n_1,n_2]_{Sp(2)} (w_1, w_2) = [n_2,n_1]_{SO(5)} (w_2, w_1)~.
\eeq

\paragraph{$SU(4)$ characters.} By a similar process as above, we find the following generating function for $SU(4)$:
\bea 
g_{SU(4)} (t_1, t_2,t_3; z_1, z_2, z_3) &=& \nn
(1 - [0,0,1] t_1t_2 - t_1 t_3 - t_2^2 - [1,0,0] t_2 t_3 \\ \nn
&+& t_1^2 t_2 + [1,0,0] t_1 t_2^2 +[0,1,0] t_1 t_2 t_3 + [0,0,1] t_2^2 t_3 + t_2 t_3^2\\ \nn
&-& t_1^2 t_2^3 - [0,0,1] t_1 t_2^2 t_3^2 - [0,1,0] t_1 t_2^3 t_3 - [1,0,0] t_1^2 t_2^2 t_3 - t_2^3 t_3^2 \\ \nn
&+& [0,0,1] t_1^2 t_2^3 t_3 + t_1^2 t_3^2 t_2^2 + t_1 t_2^4 t_3 + [1,0,0] t_1 t_2^3 t_3^2 - t_1^2 t_2^4 t_3^2 ) \\ & \times &\PE \left [ [1,0,0] t_1+ [0,1,0]  t_2 + [0,0,1] t_3 \right ]~.
\eea
The character of the irreducible representation $[n_1, n_2, n_3]_{SU(4)}$ can be simply read out from the coefficient of $t_1^{n_1} t_2^{n_2} t_3^{n_3}$ in the power series of this expression.

\end{document}